\documentclass[journal]{IEEEtran}
\usepackage[utf8]{inputenc}
\usepackage[table]{xcolor}

\usepackage{algorithm}
\usepackage{algpseudocode}
\usepackage{booktabs}
\usepackage{enumitem}

\usepackage{url}
\usepackage{cryptocode}
\usepackage{lipsum}

\usepackage{tikz}
\usetikzlibrary{calc,shapes}

\newcommand*{\rom}[1]{\expandafter\@slowromancap\romannumeral #1@}
\newcommand{\RNum}[1]{\uppercase\expandafter{\romannumeral #1\relax}}

\DeclareRobustCommand{\bitcoin}{{%
  \normalfont\sffamily
   \raisebox{-.05ex}{\makebox[.1\width][l]{-\kern-.2em-}}B %
}}

\algdef{SE}[SUBALG]{Indent}{EndIndent}{}{\algorithmicend\ }%
\algtext*{Indent}
\algtext*{EndIndent}

\ifCLASSINFOpdf
\else
\fi
\begin{document}
\title{Bitcoin Payment-Channels for Resource Limited IoT Devices}

\author{Christopher~Hannon,~\IEEEmembership{Student Member,~IEEE,}
        and Dong~Jin,~\IEEEmembership{Member,~IEEE,}% <-this % stops a space
\thanks{C. Hannon and D. Jin are with the Department
of Computer Science, Illinois Institute of Technology, Chicago,
IL, 60616 USA e-mail: channon@iit.edu, dong.jin@iit.edu.}% <-this % stops a space
\thanks{Manuscript received December 18, 2018.}}

\maketitle

\begin{abstract}
Resource-constrained devices are unable to maintain a full copy of the Bitcoin Blockchain in memory. This paper proposes a bidirectional payment channel framework for IoT devices.
This framework utilizes Bitcoin Lightning-Network-like payment channels with low processing and storage requirements. 
This protocol enables IoT devices to open and maintain payment channels with traditional Bitcoin nodes without a view of the blockchain. 
Unlike existing solutions, it does not require a trusted third party to interact with the blockchain nor does it burden the peer-to-peer network in the way SPV clients do. 
The contribution of this paper includes a secure and crypto-economically fair protocol for bidirectional Bitcoin payment channels.
In addition, we demonstrate the security and fairness of the protocol by formulating it as a game in which the equilibrium is reached when all players follow the protocol. 
%We investigate the application of our protocol in the context of Smart Grid advanced metering infrastructure billing where smart meters continuously exchange payment with a utility provider.
\end{abstract}
\IEEEpeerreviewmaketitle


\section{Introduction}
% note: because the title is kind of weird there 
\if 0
\IEEEPARstart{I}{nternet} of Things (IoT) services and devices are expanding at a rapid pace due to the expansion of networking technologies.

List of IoT:

\ 

Smart city and infrastructure,  
automobiles,  
agriculture,
sensors, 
household stuff

\ 

IoT devices paying things:

smart meters for utilities -- electricity, gas, water, internet,
cars charging, 
electronic assistants,
??

Blockchain technology has been proposed as a solution for challenges in IoT. \cite{IoT:blockchain} overviews some challenges including privacy, and value transfer while discussing how blockchain can facilitate sharing of services and the automation of work flows.

\fi

\IEEEPARstart{I}{nternet} of Things (IoT) services and devices are expanding at an exponential pace due to the rapid expansion of networking technologies. Today many companies are jumping into an IoT arms race across various application domains including smart home, connected health, wearables, connected car, smart retail,  supply chain, and many more. 
One key observation is that smart IoT devices are increasingly replacing our physical credit cards, enabling a faster and easier way for us to order products and pay services on demand. However, many problems exist concerning payment services through IoT devices, such as identity verification, security and privacy (e.g., financial information protection), scalability and flexibility (e.g., accidental ordering, refunds). 

Blockchain technology has been proposed to play a powerful role to address those challenges in IoT payment services. Blockchains are based on cryptographically secured, immutable distributed ledger technology which operate in a distributed fashion, and thus have the potential to enhance IoT solutions with better automated resource optimization, data security and reliability. For example, \cite{IoT:blockchain} describes how blockchain can facilitate sharing of services and the automation of work flows; \cite{blockchain:energy:2} overviews the blockchain integration and projects within the energy sector including markets, operations and stability, and security of the grid.

The integration of IoT and blockchain has huge potential in revolutionizing IoT. IoT devices that interact with the physical world can transfer value in exchange for services and blockchains can provide a value transfer protocol. 
Applications in Industrial IoT such as metering infrastructure in utilities including gas, electricity and water as well as electric vehicle charging and supply chain management can benefit from value transfer using blockchain technology. 
Transactions on blockchains can build trust between devices without relying on a trusted third party intermediary. 
Decentralized trustless value ledgers in the form of blockchains have gained increasing traction as trust-less value transfer protocols.
%Blockchains have caused widespread disruption across many domains including finance, insurance, internet of things, advertisement, entertainment, and even smart grid infrastructure.
Another novelty of blockchain technology is the design of a crypto-economic consensus algorithm which relaxes the assumption that some number of agents are honest to economically rational.
This creates a state that as long as participants value money (or digital cash), they will behave in a way that results in their own best interest, i.e., highest profits.
By design, blockchain consensus ensures correct operations of a decentralized public database that records users' account balances.
Blockchain technology allows for users to transfer value to other users without the help of trusted third parties such as PayPal or Visa.

%\todo{add other sectors or remove some of the specifics}

%he energy sector has seen a number of proposals designed to leverage blockchain's decentralization and trustless systems \cite{blockchain:energy:1,blockchain:energy:2,blockchain:energy:3,blockchain:energy:4,blockchain:energy:5,blockchain:energy:7,blockchain:energy:8}. 
%In \cite{blockchain:energy:2}, the authors provide an overview of blockchain integration and projects within the energy sector including markets, operations and stability and security of the grid.

%Furthermore energy companies such as BAS Nederland, Enercity, Elegant, and Marubeni all accept Bitcoin payments \cite{blockchain:energy:9}. companies such as PowerLedger (\url{https://powerledger.io/}), and LO3 Energy (\url{https://lo3energy.com/}), offer peer-to-peer renewable energy marketplaces. 

\begin{figure}
\centering
\includegraphics[scale=.625]{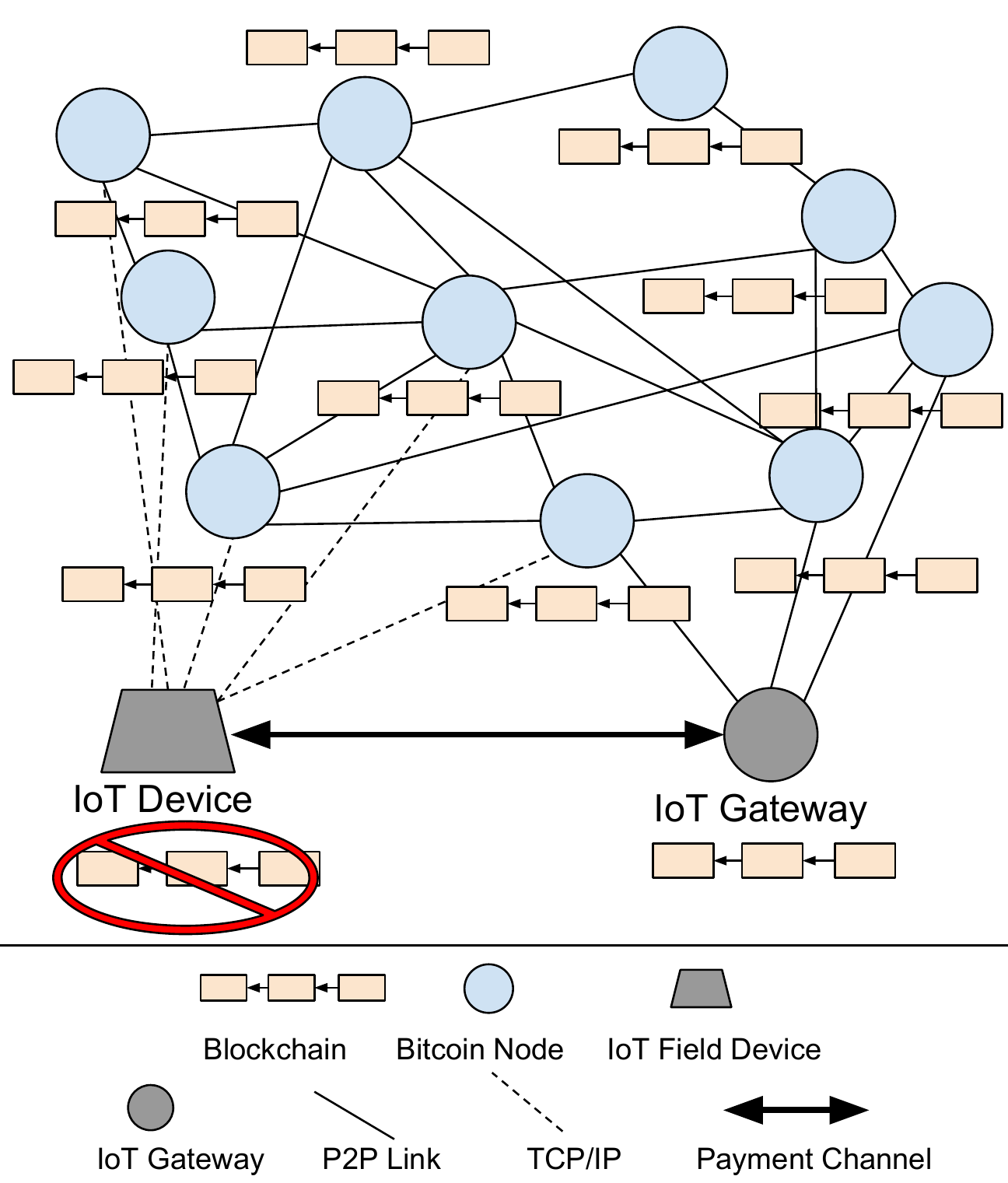}
\caption{IoT devices in the Bitcoin Blockchain context. The IoT device has a payment channel open to a gateway service and does not have a local copy of the blockchain. Instead, the IoT device relies on untrusted third parties to connect on its behalf through economic incentives.}
\label{iot_overview}
\end{figure}

The main advantage of blockchain is its trust-less value transfer protocol, which is securely maintained through decentralized participants.
However, the blockchain technology does not solve all problems, specifically, blockchains suffer from limited scalability due to their decentralized nature.
Additionally, the limited scalability can drive up the cost of using the blockchain network through high fees.
In this work, we focus on using blockchain payment channels which enables scalability. We design a payment channel protocol based on the Lightning Network, which enables IoT devices with few computational and storage resources to transfer value.
In particular, our protocol enables a party to transact with another using the Bitcoin blockchain without storing the complete blockchain using untrusted third parties. Figure~\ref{iot_overview} shows the architecture of our blockchain protocol in reference to the rest of the blockchain. 

The remaining paper is organized as follows: Section~\ref{background} provides background on blockchain and its payment channels. %, and IoT and cyber physical systems. 
Section~\ref{protocol} presents a protocol design that enables real-time payment channels for IoT devices to gateway services. We analyze the security of the protocol by formulating it as a game in Section~\ref{analysis}. 
Finally, we describe the related work in Section~\ref{related-work} and conclude in Section~\ref{conclusion} with future work directions.

\label{introduction}

\section{Background}
\label{background}
Blockchain technology at its core is an immutable public digital ledger containing transactions.
The novelty of blockchain is its ability to unequivocally agree on the state of the ledger in a decentralized setting. 
Through the process of \textit{mining}, the global state of the blockchain advances. 

The Bitcoin blockchain \cite{BTC:whitepaper} provides the ability to send transactions which consist of inputs, outputs, and rules governing the redemption of the outputs.
Inputs to transactions map to the source of the funds (a previous transaction) while the outputs represent the destination of the transaction value.
The governing rules included in the transaction dictate how the recipient of the transaction is able to spend the received value in the future.
A common rule is for the recipient to prove ownership of a private key associated with the destination address in the transaction. 
However, more detailed rules can be expressed to provide more complicated value redemption logic.

In general, Bitcoin follows the UTXO model which says that an input to a new Bitcoin transaction is the output of an unspent previous transaction along with a script that redeems the previous transactions output. The output of the new transaction provides the recipient and a script telling the recipient how to redeem the values. 
Since the blockchain is an immutable public database, transactions cannot be revoked and the total of all unspent transaction outputs represent the current state of the Blockchain.
Thus, there is no concept of users or accounts included in the Bitcoin Blockchain.

Transactions are organized into blocks which remain pending until a partial pre-image is found for the sha-256 
%\footnote{In implementation the pre-image is hashed twice with sha-256.} 
 hash algorithm which meets a specific criteria quantified as the blockchain's difficulty. 
This difficulty is a dynamic variable that corresponds to the processing power of participants working to add new blocks and transactions into the blockchain. 
The result of this process maintains that on average new blocks are added to the blockchain every 10 minutes.
In alternative blockchain implementations, the target block interval varies, e.g., 15 seconds in Ethereum \cite{ETH:whitepaper}.
This interval is important to note because until a transaction is included in a block it is not considered verified by the blockchain network.
Furthermore, due to the consensus algorithm that governs the blockchain, there may be a temporary fork where multiple valid blocks are at the same height.
A block is only valid if it is part of the longest blockchain.
Confidence of immutability grows exponentially in relation to the depth of the block, i.e., number of subsequent blocks. 
The original Bitcoin white paper \cite{BTC:whitepaper} provides a more detailed analysis. However, one heuristic used in practice is 6 blocks (about 60 minutes)\cite{BTC:wiki}.

\begin{figure}
\centering
\includegraphics[scale=.44]{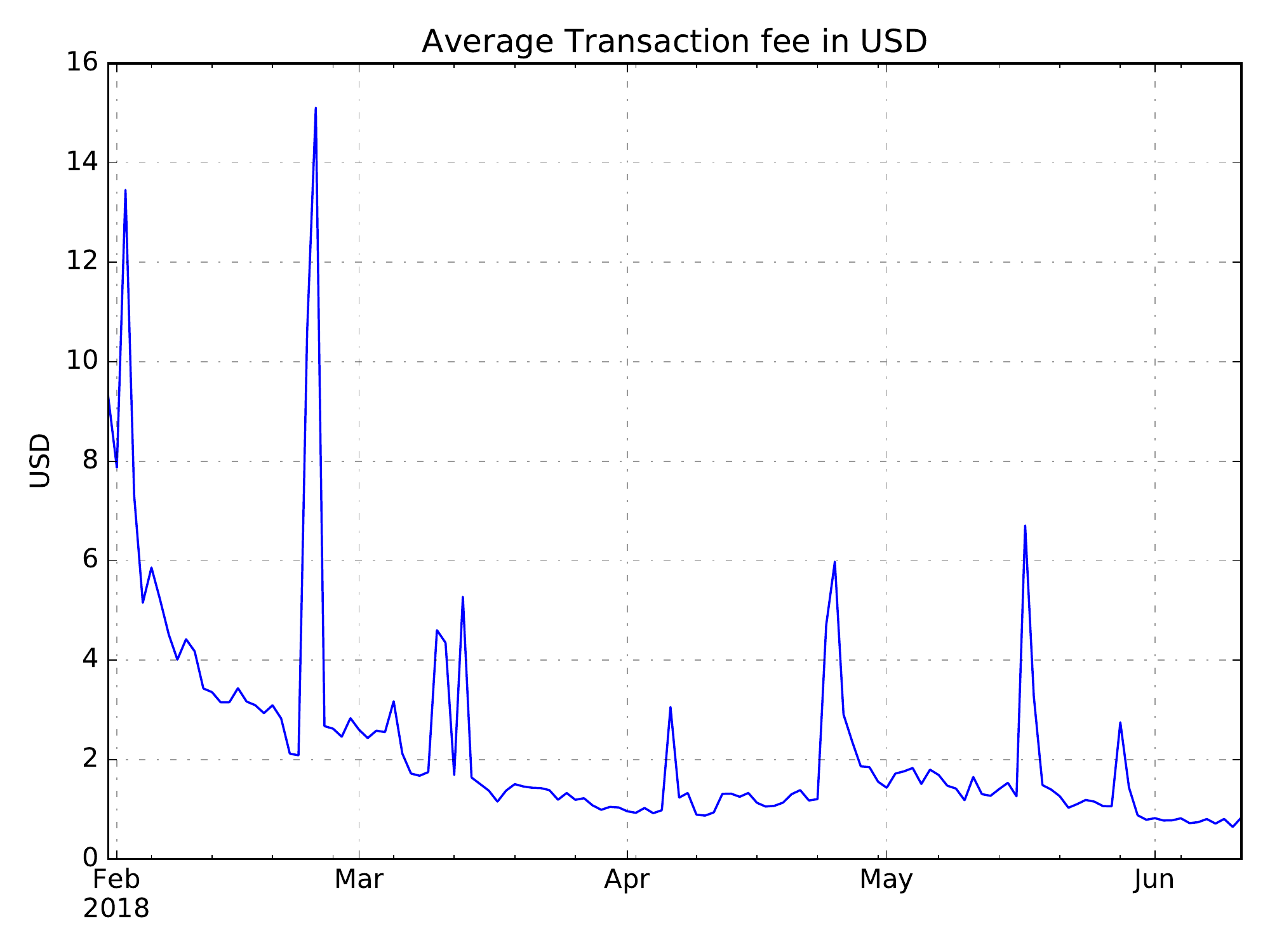}
\caption{Average transaction fee from Feb. to June 2018 in USD. Calculated using the blocksci library \cite{btc:block:sci}.}
\label{fees}
\end{figure}

Figure~\ref{fees} shows the cost of sending a transaction converted to USD over four months in 2018. 
The cost of a transaction on the blockchain as well as the time required to publish the transaction makes frequent real-time transactions impossible. %, a single transaction per month does not incur outrageous fees. 
%However that is equivalent to the existing billing strategy.
Another method to reduce fees is to use an alternative blockchain that has larger block sizes or more frequent blocks.
Although alternative blockchains can have weaker security, and greater price volatility that 
%A single monthly transaction is closer to the existing billing solutions which 
does not satisfy our goals in this work.
In this paper, we propose a protocol using \textit{off-chain} payment channels that can provide \textit{real-time} payment but do not incur large fees with frequent posting to the blockchain.

\subsection{Off-Chain Bitcoin Transactions}
Bitcoin's Forth-like scripting language enables more complex functionality by placing conditions on the redeeming of transaction outputs. 
For example, time locks can be used on transactions and transaction outputs to place temporal restrictions on the ability to spend or create transactions.
Time locks can be both relative and absolute and can prevent a transaction output from being spent until after a certain time. 
Time locks are one of the most important building blocks in off-chain Bitcoin Transactions, such as in the Bitcoin Lightning Network \cite{BTC:lightning}.
Multisignatures can require multiple keys to spend a transaction output.
One use case for multisignatures is joint savings accounts where both parties need to agree to make a transaction. 
%Multisignatures
%A Hashed TimeLock Contract (HTLC) is a type of contract that builds off timelock and hashlock scripting primitives. 
%HTLCs are contracts that require additional information in order to make the transaction valid. 
%The timelock requires a temporal attribute to become valid while the hashlock requires a hash function pre-image to validate a transaction.
Time lock contracts include primitives \cite{BTC:wiki} such as
\begin{itemize}
    \item \texttt{nLockTime} specifies the minimum height of the blockchain that a transaction can be included in.
    \item \texttt{CheckLockTimeVerify} requires that the blockchain be at a certain height for the output of an already included transaction to be spent.
    \item \texttt{Relative LockTime} places restrictions on the inclusion of an input to a new transaction based on the time that the input was included in the previous transaction. %i.e., restricting the time sent BTC can be sent again.
    \item \texttt{CheckSequenceVerify} provides a relative time that the output of a transaction becomes valid after inclusion in a block. 
    \item \texttt{Multisig} can be used to require $m-\text{of}-n$ signatures to become valid.
\end{itemize}
\texttt{nLockTime} and \texttt{Relative Locktime} are corresponding counterparts for absolute and relative time respectively for inclusion into the blockchain while  \texttt{CheckLockTimeVerify} and \texttt{CheckSequenceVerify} are counterparts for absolute and relative time respectively for making outputs of a valid transaction spendable.

Hashlocks on the other-hand provide encumbrance on the outputs of transaction that requires a specified secret value being publicly revealed.
%that require as input some secret value that is the pre-image to some hash functions output.
Upon unlocking the hashlock, all other hashlocks with the same secret value are also unlocked due to the secret value being recorded on the blockchain.

The combination of hashlocks and timelocks can create timed hashlock contracts (HTLCs) which can be used to put Bitcoin transactions 'off-chain' through what is called L2 or layer 2 scaling solutions, such as the Lightning Network \cite{BTC:lightning}, Duplex micropayment channels\cite{BTC:duplex}, and Raiden \cite{ETH:Raiden}. 
Our protocol uses \texttt{CheckSequenceVerify} and \texttt{Multisig} for payment channels similar to Lightning Network Transactions.
Off-chain transactions are properly formatted Bitcoin transactions that are deferred from being published immediately on the blockchain. 
The benefit is that off-chain transactions can be updated many times before published to the blockchain resulting in reduced on-chain transactions and ultimately fewer fees.
In \cite{BTC:duplex}, channels can be created unidirectional or in duplex, and transactions can be updated so that the final channel balances are guaranteed to be included. 
This is accomplished by newer transactions having smaller timelocks than earlier transactions, and thus being able to be published sooner than old invalidated transactions. 
A limitation of this approach is that channels will have finite lives.
References \cite{BTC:lightning} and \cite{ETH:Raiden} allow for channels to remain open indefinitely or until the owners decide to close the channel. 
%As distributed energy resources such as solar, storage technology, and wind power become more prevalent in the modern electric power grid, it is possible for end users to produce more energy than they consume at various times of the day.
%Therefore a payment protocol for AMI should allow for bidirectional payments. 

In those solutions, transactions are properly formatted so that either party involved can post the transaction at any time to the blockchain in case of dispute.
This property allows for the protocol to remain trustless. To avoid the problem of excessive fees, transactions are updated off-chain to reflect a new balance and only publish them upon closure to the blockchain.
Therefore, fees only need to be paid when opening and closing a channel and updating the balance within the channel is fee-less.

%The major challenge in off-chain transactions is that when updating the channel balance by creating a new transaction, the old unpublished transactions need to be invalidated. 
%The Lightning Network \cite{BTC:lightning} addresses this issue by using Revocable Delivery Transactions, which are designed using timelocks and Bitcoin Scripts to enable users to forfeit funds once they publish old channel states to the blockchain.
%Correct behavior is ensured through the assumption that actors are economically rational.
%The Lightning Network ensures that if an actor misbehaves they can not end up any better off than if they follow the protocol.
%The natural limitation to the Lightning Network is that the channel participants need to have access to the blockchain to follow the protocol correctly. 
%If one party does not have access to the blockchain, the counter-party may be able to steal funds by publishing old channel states. 
%The parties need to watch the blockchain to ensure that if an old state is published that they take appropriate actions.

Because IoT devices are resource limited, it is infeasible to assume that they can remain connected to the blockchain. 
Therefore, we need to adopt the state channel model to allow for one (or both) parties to be offline from the Bitcoin blockchain.
In this work, we assume that the IoT devices have networking capability to untrusted third parties. 
By providing financial incentives to the third parties, the IoT devices do not need to have direct access to the blockchain and can instead rely on the untrusted third parties to act as a bridge to ensure correct operation of the protocol.

\section{Protocol Design}
\label{protocol}
%\todo{make section general for IoT, make it more formal--}

%Definitions of the variables used in the following procedures and protocol:
%\begin{itemize}
%\item $ t_C \text{ : The interval that data is collected at}$
%\item $ t_R \text{ : The interval that data is reported at} $
%\item $ n = \frac{t_R}{t_C}  \text{ : The number of measurements / report interval} $
%\item $ \Vec{p} = \left<p_1,p_2,...,p_n \right>$ Vector of n prices\footnote{\bitcoin \ is the unit of value in the Bitcoin blockchain that can be broken down into $10^9$ smaller units. 1 \bitcoin \ $ = 10^9$ Satoshis } in \bitcoin$  \e{-9}$
%\item $ \Vec{m} = \left<m_1,m_2,...,m_n \right> \text{Vector of $n$ measurements}$
%\item $T_{FT}$ : The funding transaction
%\item $T_{C}$ : A Commitment transaction
%\item $in_k$ : A input in a Bitcoin transaction
%\item $out_k$ : A output in a Bitcoin transaction
%\item $\sigma$ : A value in Satoshi ($10^{-9} \times $\bitcoin)
%\item $(pk,sk)$ : A keypair 
%\item $tx\_id$ : Some valid transaction ID and output index
%\item $ rd $ : A redemption script whƒich requires $rd'$ to make valid
%\item $A$ : 'Alice' -- an IoT device with limited resources
%\item $B$ : 'Bob' -- an IoT gateway device using in payment channel
%\item $j$ : The number of intermediate channel states (number of keys)
%\item $a|b|c$ : denotes 3 posible 
%\item $(pk_{\_\{B|A\}\_j\_\{a|b|c\}},sk_{\_\{B|A\}\_j\_\{a|b|c\}})$ -- a keypair for $A$ or $B$
%\item $3rd$ : A third party
%\end{itemize}

The intuition of our protocol is to use a multisig transaction to fund the channel.
Intermediate states are made revocable as developed in \cite{BTC:lightning}.
Our contribution is to ensure correctness and crypto-economical fairness when one party does not have access to the blockchain.
To do this, we use a third party to post transactions to the blockchain by creating an additional output that is spendable by the third party. 
By creating an economic incentive, a third party is willing to participate in the protocol. 
Furthermore, to ensure that the second party does not violate the protocol by publishing a revoked state, a third party is used as a \textit{watchdog}, which informs the first party when the funding transaction's output is used as an input to a new transaction.
This \textit{watchdog} will report when an intermediate state is posted to the blockchain, which prevents publishing an expired state.
Additionally, to prevent collusion between the second party and the third parties, we use a pool of third parties for each service. 
Since any member in the third party is able to take the role, the incentive needs to be higher than any incentive from colluding. 
In Section~\ref{analysis}, we formulate the problem as a game and show that the equilibrium is reached by following the protocol.

The details of the protocol are as follows. For each IoT device $A$, the IoT payment gateway $B$ creates a payment channel. 
Both $A$ and $B$ generate 3 sets of $j$ keypairs $(pk_{\_\{B|A\}\_j\_\{a|b|c\}},sk_{\_\{B|A\}\_j\_\{a|b|c\}})$ such that each intermediate transaction %, state of payment channel, 
uses a different keypair making $j$ the number of intermediate states. 
Additionally, we generate a keypair for opening and closing the channel, $(pk_{\_\{B|A\}\_\{FT|close\}},sk_{\_\{B|A\}\_\{FT|close\}})$, and another pair for $A$ transacting with the third parties, $(pk_{\_A\_3rd\_\{a|rc\}},sk_{\_A\_3rd\_\{a|rc\}})$.
To minimizes the IoT devices' memory requirements, we use BIP32 that 
provides a deterministic hierarchical key generation algorithm with a highly compacted data structure \cite{BTC:bip32}.
All keys can then be generated with a given master key and an index in the data structure. 
Effectively this enables the storing of an index in place of a keypair, as the total requirements for storing a state is the key index and the balances. 
%The specific transactions can all be generated from this information.

The two parties agree to place a Funding Transaction $T_{FT}$ on the blockchain that sends $\Omega_{A}$ and $\Omega_{B}$ as input funds from $A$ and $B$ respectively. 
The output of the channel is a 2-of-2 multi-signature requiring both $A$ and $B$'s $sk_{\{B|A\}\_FT}$ to spend. 
Additionally, the two parties agree on an initial commitment transaction $T_{C1}$ that is a valid spending of the funds from $T_{FT}$ as the input and returning $\Omega_{A}$ and $\Omega_{B}$ as the outputs. 
Note that the transaction is not published to the blockchain, and its purpose is to denote the starting balance in the payment channel. 
Reference \cite{BTC:lightning} shows how two transactions, $T_{C1\_B}$ and $T_{C1\_A}$ can be constructed so that $B$ is the only party that can publish $T_{C1\_A}$ and $A$ is the only party that can publish $T_{C1\_B}$. 
This mechanism is accomplished by supplying one of the 2-of-2 input signatures required to spend $T_{FT}$'s output, i.e., partially signing the transaction.

Furthermore, the transactions $T_{C1\_B}$ and $T_{C1\_A}$ are made revocable by encumbering the outputs of the corresponding party's ability to send the output. 
For example, if $A$ publishes $T_{C1\_B}$, the output of $\Omega_{B}$ can be redeemed immediately by $B$. However, $\Omega_{A}$ funds are locked.
There are two ways to redeem the locked funds. The first is to use $A$ and $B$'s secret keys $sk_1$s for a 2-of-2 multi-signature.
The second is to use a timelock to redeem the fund to $A$ after $W$ blocks, where $W$ is the number of blocks specified in the timelock. 
 $B$ may use the first mechanism to \textit{steal} all of $A$s funds after both parties update the state of the channel to the new transactions $T_{C2\_B}$ and $T_{C2\_A}$. Upon updating, $A$ sends $sk_1$ to $B$ (and $B$ sends $sk_1$ to $A$).
The \textit{stealing} of funds relies on the mechanism to prevent \textit{old} transactions from being published, which can be achieved by $B$ checking the blockchain before some $W$ blocks after $T_{C1\_B}$ is published.
This mechanism is a way to ensure that both parties follow the protocol even if the two parties do not trust each other. 

However, since the IoT devices are assumed not to have a direct access to the blockchain, two disjoint groups of untrusted third parties are used to interface between the IoT device and the blockchain.
Each group has multiple members, $K_1$ and $K_2$ respectfully, to prevent collusion with $B$.
The first group is used to publish a transaction to the blockchain incentivized through a small fee as an output to the transaction.
The second group ensures that if $B$ publishes an old transaction that the IoT device is notified of the transaction and is able to spend the transaction output before the timelock $W$ expires for $B$ to redeem their funds. 
This second group is also incentivized through small fees in Bitcoin smart contracts.  Transactions~\ref{commitmentTXa}-\ref{RecoverTX} show the method to do this.
In order to prevent third parties colluding with each other or  $B$, the number of members in each third party has to be chosen with respect to the quantity of fees as well as to the channel balances.
%We show in the following section how this is done.

When both parties agree on the closing of a channel, they can create a new transaction $T_{Fin}$ that uses $\Omega_{A\_fin}$ and $\Omega_{B\_fin}$
as the final output balances, and post it to the blockchain.
If both parties follow the protocol properly, only two transactions, $T_{FT}$ and $T_{Fin}$, are published to the blockchain.
If one party tries to publish an old state of the channel $T_{Ci\_(B/A)}$, the other party can detect this and take all the funds in the channel. 

Transaction~\ref{FundTX}, the funding transaction, contains two or more inputs, and one or more outputs. 
The channel funding output is a multisignature output requiring both parties to sign in order to use as an input into a new transaction.

Transaction~\ref{CloseTX}, is used upon mutual channel closing, it uses the multisignature output from the funding transaction and uses multiple outputs to both $A$ and $B$. 
If the IoT device wishes to post the transaction, a fee $\sigma$ can be placed in an input for a third party to be incentivized to publish.

Transaction~\ref{commitmentTXa} takes the funding transactions output as input and creates 3 outputs. 
The first output is local to A, the party with the ability to publish it.
This output is encumbered by a timelock of $W$ blocks to $A$'s address to ensure that if the transaction is old. In other words, A has given a key pair $(pk_{\_\{A\}\_j\_\{c\}},sk_{\_\{A\}\_j\_\{c\}})$ to B, and B can redeem this input.
The second output is the remote output to $B$.
Finally, there is a third output, which is the incentive for a third party to send the transaction to the blockchain and make sure the transaction gets included in a block.

Transaction~\ref{commitmentTXb} takes the funding transactions output as input and creates 2 outputs. 
Because $A$ partially signs the input to this transaction, only $B$ is able to publish it.
The first output is timelocked by $W$ blocks with an output to $B$. 
$A$ is also able to redeem this input given $(pk_{\_\{B\}\_j\_\{c\}},sk_{\_\{B\}\_j\_\{c\}})$. 
There is a third party watching the blockchain, whose key is required for $A$ to redeem this input. By doing so, the third party also gets a fee in return.

Transaction~\ref{RecoverTX} takes this as input and can be presigned. This recovery transaction is used as a smart contract to incentivize a third party to watch the blockchain by providing a fee determined by the members of the third party.
The second output to Transaction~\ref{commitmentTXb} is used for a remote output to $A$.

The Bitcoin Scripts used for all the aforementioned transactions are shown in Appendix~\ref{bitcoinScripts} to provide a real-time payment channel for IoT devices with an IoT gateway.

%The formal protocol is shown in Figure~\ref{protocol_Fig} and the Bitcoin Scripts used are shown in Appendix~\ref{bitcoinScripts} to provide a real-time payment channel for IoT devices with a IoT gateway.

\section{Security Analysis}
\label{analysis}
To prove that our protocol design is crypto-economically fair, we model the protocol as a game and demonstrate that the equilibrium can always be reached as long as the players follow the protocol and fees are appropriately set. 

The payment channel in the Bitcoin Lightning Network \cite{BTC:lightning} can be modeled as a game between two actors. 
After the channel is funded, Player I may post any previous states and then Player II may choose to follow the protocol or deviate. 
Following the protocol means to take the maximum amount of funds, i.e., the remote transaction as well as the local transaction, if the transaction is rescinded. 
Deviating means to do nothing or to take just the remote funds.
Let us consider 3 transactions with Players I and II balances $\alpha$, $\beta$ respectively. We define $TX_1 = (\alpha_1,\beta_1)$, $TX_2 = (\alpha_2,\beta_2)$, and $TX_3 = (\alpha_3,\beta_3)$, such that $\alpha_2 > \alpha_1 > \alpha_3$ and $\beta_3 > \beta_1 > \beta_2$, where $TX_1$ is the current state of the channel, and $TX_2$ and $TX_3$ are previous states where $\alpha$ and $\beta$ are the values each party has respectively in the channel at a state. 
Additionally, $\alpha_1+\beta_1 = \alpha_2+\beta_2 = \alpha_3+\beta_3$ since the total amount of funds in the channel is fixed.
Player I's strategies are which TX to publish to the blockchain. 
Following the protocol, the strategy is publishing $TX_1$, while $TX_2$ and $TX_3$ is deviating from the protocol.
Player II's strategies are Follow, Deviate\_1, and Deviate\_2 as described above.
The payoff matrix for this game is shown in Table~\ref{ln_mx}. Player I experiences a maximum payout under strategy D\_1 if Player II chooses a deviating strategy. However, since Player II has a pure strategy always to follow the protocol, the equilibrium is reached when Player I also follows the protocol. 
In this game, we show that in the life of the channel, no player will be able to increase their profit if the other player follows the protocol. 

\begin{table}
    \centering
    \begin{tabular}{ c|c c c } 
        I/II & F & D\_1 & D\_2 \\
        \hline
        %\vspace{1mm}
        \rowcolor{lightgray} F   & \cellcolor{gray} $\beta_1$ / $\alpha_1$ & $\alpha_2+\beta_2$ / $0$ & $\alpha_3+\beta_3$ / $0$ \\ 
        D\_1 & $\beta_1$ / $\alpha_1$ & $\beta_2$ / $\alpha_2$ & $\beta_3$ / $\alpha_3$ \\ 
        D\_2 & $0$ / $\alpha_1$ & 0 / $\alpha_2$ & 0 / $\alpha_3$ \\ 
    \multicolumn{4}{c}{\ } \\
    \end{tabular} 
    \caption{ Payoff matrix of payment channel game between 2 parties}
    \label{ln_mx}
\end{table}

In our approach, we assume that one party does not have access to the blockchain, and also with the addition of more players, the game gets further complicated. 
%With the addition of more players, the protocol that we model as a game becomes more complicated.
The interesting cases to evaluate are when one of the parties posts an intermediate state to the blockchain. 
Let us consider four players in the game, the IoT gateway, an IoT device, and two groups of untrusted third parties. 
Player 1 is the IoT gateway with three strategies, using the same set of transactions as the previous game. 
Each strategy refers to posting a transaction to the blockchain where strategy 1 is following the protocol.
In the first game where player 1 goes first (see Figure \RNum{3}), %~\ref{expansive-form-1}),
%Figure~\ref{expansive-form-2} 
%Figure~\ref{sizes}  
%\todo{This is reporting wrong Figure}
Player 4, the 3rd party that watches the blockchain for an output of the funding transaction in a new transaction plays next.
Player 4 can either tell player 2 about the transaction (Strategy F) or deviate from the protocol. 
Since Player 4 represents a group of players, $K_2$, any one of them can follow the protocol. 
Therefore, in order to deviate, all Players in the group must collude. 
If the players collude to perform a denial-of-service attack, then no profit is gained. Therefore, it is not rational.
On the other hand, if the Player 4 members collude with Player 1 (Strategy D), then they can receive some payoff $\frac{\gamma_2}{K_2}$, where $K_2$ is the number of members in the group and $\gamma_2$ is the amount that Player 1 offers which is bounded by $\alpha_1$.

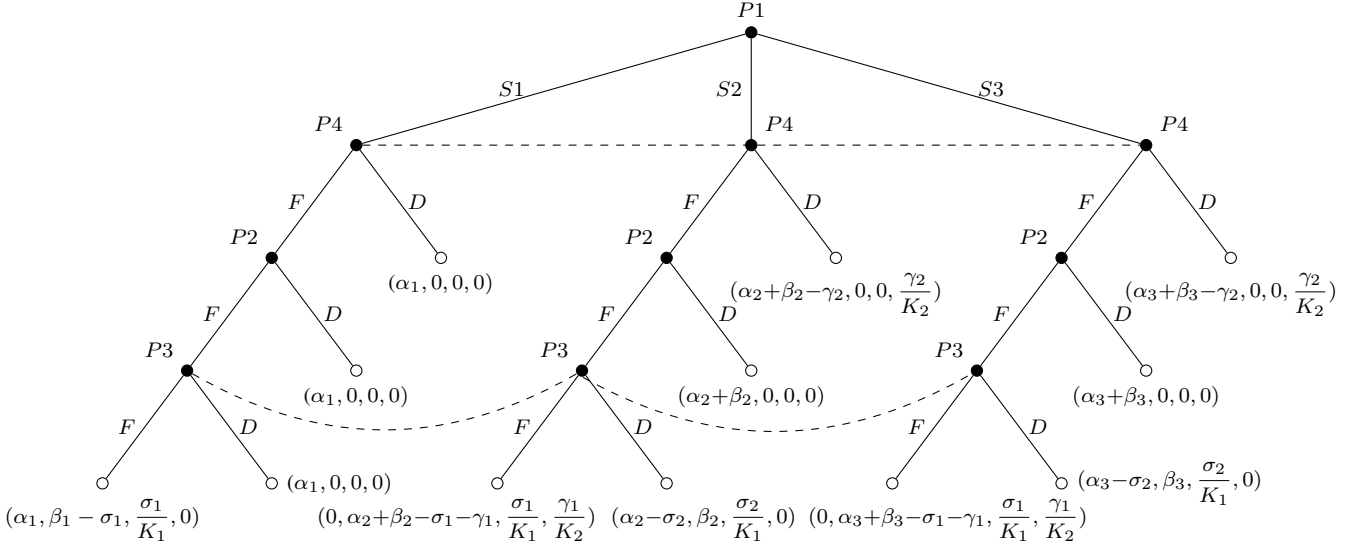
\begin{figure*}
\centering
    \label{expansive-form-1}
    \begin{tikzpicture}[scale=1.5,font=\footnotesize]
    \tikzstyle{solid node}=[circle,draw,inner sep=1.5,fill=black]
    \tikzstyle{hollow node}=[circle,draw,inner sep=1.5]
    \tikzstyle{dashe}=[edge,draw,dashed]
    \tikzstyle{level 1}=[level distance=10mm,sibling distance=3.5cm]
    \tikzstyle{level 2}=[level distance=10mm,sibling distance=1.5cm]
    \tikzstyle{level 3}=[level distance=10mm,sibling distance=1.5cm]
    \tikzstyle{level 4}=[level distance=10mm,sibling distance=1.5cm]
    \tikzstyle{level 5}=[level distance=10mm,sibling distance=2cm]
    \node(0)[solid node,label=above:{$P1$}]{}
        child{node[solid node,label=above left:{$P4$}](s1){}
            child{node[solid node,label=above left:{$P2$}]{} 
                child{node[solid node,label=above left:{$P3$}](a){} 
                    child{node[hollow node,label=below:{$(\alpha_1,\beta_1-\sigma_1,\dfrac{\sigma_1}{K_1},0)$}]{} 
                        edge from parent node[left]{$F$}
                    }
                    child{node[hollow node,label=right:{($\alpha_1,0,0,0)$}]{} 
                        edge from parent node[right]{$D$}
                    }
                edge from parent node[left]{$F$}
                }
                child{node[hollow node,label=below:{($\alpha_1,0,0,0)$}]{} 
                    edge from parent node[right]{$D$}
                }
                edge from parent node[left]{$F$}
            }
            child{node[hollow node,label=below:{($\alpha_1,0,0,0)$}]{} 
            %child{node[hollow node,label=below:{($\alpha_1{-}\gamma_2,0,0,\dfrac{\gamma_2}{K_2})$}]{} 
               edge from parent node[right]{$D$}
            }
            edge from parent node[left,xshift=-7.5]{$S1$}
        }
        child{node[solid node,label=above right:{$P4$}](s2){}
            child{node[solid node,label=above left:{$P2$}]{} 
                child{node[solid node,label=above left:{$P3$}](b){} 
                    child{node[hollow node,label=below:{$(0,\alpha_2{+} \beta_2{-}\sigma_1{-}\gamma_1, \dfrac{\sigma_1}{K_1},\dfrac{\gamma_1}{K_2})\ \ \ \ \ \ \ \ \ \ \ $}]{} 
                        edge from parent node[left]{$F$}
                    }
                    child{node[hollow node,label=below:{$\ \ \ \ \ \ \ \ \ \ (\alpha_2{-}\sigma_2,\beta_2,\dfrac{\sigma_2}{K_1},0)$}]{} 
                        edge from parent node[right]{$D$}
                    }
                edge from parent node[left]{$F$}
                }
                child{node[hollow node,label=below:{$(\alpha_2{+}\beta_2,0,0,0)$}]{} 
                    edge from parent node[right]{$D$}
                }
                edge from parent node[left]{$F$}
            }
            child{node[hollow node,label=below:{$(\alpha_2{+}\beta_2{-}\gamma_2,0,0,\dfrac{\gamma_2}{K_2})$}]{} 
                edge from parent node[right]{$D$}
            }
            edge from parent node[left,]{$S2$}
        }
        child{node[solid node,label=above right:{$P4$}](s3){}
            child{node[solid node,label=above left:{$P2$}]{} 
                child{node[solid node,label=above left:{$P3$}](c){} 
                    child{node[hollow node,label=below:{$\ \ \ \ \ \ \ \  \ \ \ \ \ \ \ (0,\alpha_3{+} \beta_3{-}\sigma_1{-}\gamma_1, \dfrac{\sigma_1}{K_1},\dfrac{\gamma_1}{K_2})$}]{} 
                        edge from parent node[left]{$F$}
                    }
                    child{node[hollow node,label=right:{$(\alpha_3{-}\sigma_2,\beta_3,\dfrac{\sigma_2}{K_1},0)$}]{} 
                        edge from parent node[right]{$D$}
                    }
                edge from parent node[left]{$F$}
                }
                child{node[hollow node,label=below:{$(\alpha_3{+}\beta_3,0,0,0)$}]{} 
                    edge from parent node[right]{$D$}
                }
                edge from parent node[left]{$F$}
            }
            child{node[hollow node,label=below:{$(\alpha_3{+}\beta_3{-}\gamma_2,0,0,\dfrac{\gamma_2}{K_2})$}]{} 
                edge from parent node[right]{$D$}
            }
            edge from parent node[right,xshift=7.5]{$S3$}
            %(s1) edge node[dashed,right]{} (s2)
            %(s2) edge node[right]{} (s3)
        };
        % information set
        \draw[dashed] (s1)to(s2)to(s3);  
        \draw[dashed,bend right] (a)to(b)to(c);  
        %\draw[dashed](0-1)to(0-2)to(0-3)
    \end{tikzpicture}
    \caption{Extensive form when Player 1 goes first. The optimal solution is reached when all players follow the protocol.}
\end{figure*}

If we show that $\frac{\gamma_2}{K_2}$ is less than $\gamma_1$, then Player 4 will not be incentivized to deviate from the protocol. 
Player 2 will not deviate from the protocol because they have a pure strategy to follow the protocol.
Finally, similarly to Player 4, Player 3 can follow the protocol where any one member of $K_1$ will earn $\sigma_1$. 
After many games, the average payout is $\frac{\sigma_1}{K_1}$. 
$\sigma_1$ must be large enough to cover the fee of spending the transaction as well as for the bandwidth requirements in order to maintain a connection to the IoT device. 
Additionally, in order to prevent collusion between all the members in Player 3 and Player 1, $\frac{\sigma_2}{K_1}$ is less than $\sigma_1$. 
This forced inequality is the reason for using multiple members in each group. 
If $\frac{\sigma_2}{K_1} < \sigma_1$ and $\frac{\gamma_2}{K_2} < \gamma_1$ are true, then the equilibrium is reached when all players follow the protocol correctly.
In order to ensure these inequalities hold, a minimum/maximum channel balance must be enforced. 
Recall that $\alpha_2 > \alpha_1 > \alpha_3$,  to evaluate the fees of $\sigma$ and $\gamma$, we can set $TX_2$ to the intermediate state where $\alpha_2 = \forall i \ MAX(\alpha_i) $ and similarly $TX_3$, $\alpha_3 = \forall i \ MIN(\alpha_i) $

Player 1 has the potential to maximize their potential earnings when they deviate with strategy 2, by posting $TX_2$ to the blockchain with earnings of $\alpha_2 - \sigma_2$ and $\alpha_2 - \gamma_2$ and by colluding with Players 3 and 4 respectively. 
By enforcing that 
$$\sigma_1 > \frac{\sigma_2}{K_1} \equiv \sigma_1 > \frac{\alpha_2 - \alpha_3}{K_1}$$ 
and similarly for Player 4, 
$$\gamma_1 > \frac{\gamma_2}{K_2} \equiv \gamma_1 > \frac{\alpha_2 - \alpha_3}{K_2}$$ 
we can show that the equilibrium is met when all players follow the protocol because Player 3 and Player 4 will not collude with Player 1.

If Player 2 makes the first move by posting a transaction to the blockchain, the game is similar to the original payment channel game shown in Table~\ref{ln_mx}. 
Player 1 will have a pure strategy to follow the protocol. 
However, Player 3 can deviate from the protocol by performing a denial-of-service attack against Player 2.
If Player 3 does this through collusion, the game actually restarts.
By incentivizing Player 3 and because there is a pool of members in Player 3, it is not economically rational to take that strategy. 
Therefore, using the fee structure for $\sigma$ as in the game when Player 1 goes first, equilibrium is reached when all players follow the protocol. 
Although in both equilibrium cases Player 4 does not get an incentive, they do not know which strategy Player 1 has taken, thus their profit is still maximized when following the protocol.

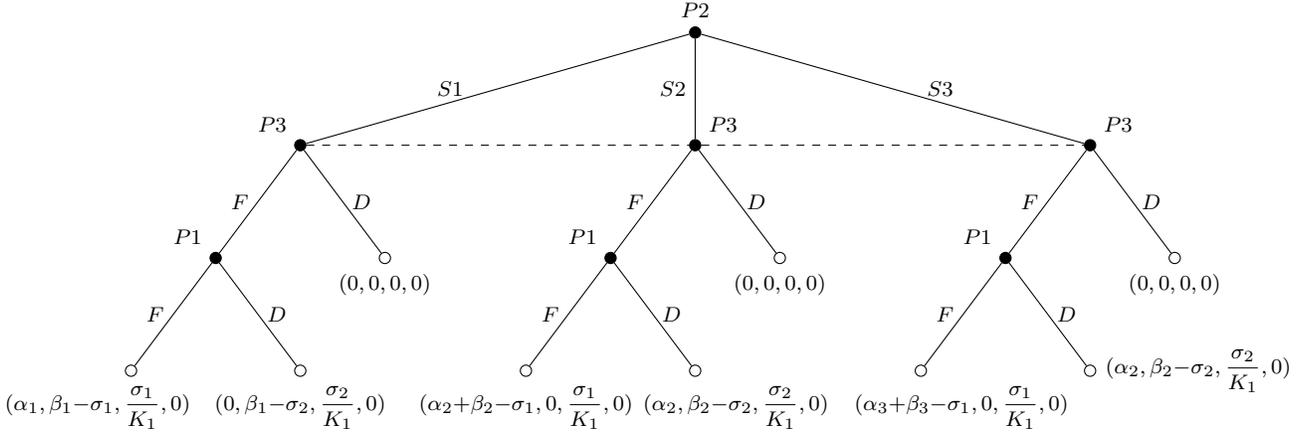
\begin{figure*}
\centering
    \label{expansive-form-2}
    \begin{tikzpicture}[scale=1.5,font=\footnotesize]
    \tikzstyle{solid node}=[circle,draw,inner sep=1.5,fill=black]
    \tikzstyle{hollow node}=[circle,draw,inner sep=1.5]
    \tikzstyle{level 1}=[level distance=10mm,sibling distance=3.5cm]
    \tikzstyle{level 2}=[level distance=10mm,sibling distance=1.5cm]
    \tikzstyle{level 3}=[level distance=10mm,sibling distance=1.5cm]
    \tikzstyle{level 4}=[level distance=10mm,sibling distance=1.5cm]
    \tikzstyle{level 5}=[level distance=10mm,sibling distance=2cm]
    \node(0)[solid node,label=above:{$P2$}]{}
        child{node[solid node,label=above left:{$P3$}](s1){}
            child{node[solid node,label=above left:{$P1$}]{} 
                    child{node[hollow node,label=below:{$(\alpha_1,\beta_1{-}\sigma_1,\dfrac{\sigma_1}{K_1},0) \ \ \ \ \ \ \ \ \ $}]{} 
                        edge from parent node[left]{$F$}
                    }
                    child{node[hollow node,label=below:{$(0,\beta_1{-}\sigma_2,\dfrac{\sigma_2}{K_1},0)$}]{} 
                        edge from parent node[right]{$D$}
                    }
                edge from parent node[left]{$F$}
            }
            child{node[hollow node,label=below:{$(0,0,0,0)$}]{} 
                edge from parent node[right]{$D$}
            }
            edge from parent node[left,xshift=-9.5]{$S1$}
        }
        child{node[solid node,label=above right:{$P3$}](s2){}
            child{node[solid node,label=above left:{$P1$}]{} 
                    child{node[hollow node,label=below:{$(\alpha_2{+} \beta_2{-}\sigma_1,0, \dfrac{\sigma_1}{K_1},0)$}]{} 
                        edge from parent node[left]{$F$}
                    }
                    child{node[hollow node,label=below:{$\ \ \ \ \ \ \ \ \ \ \ (\alpha_2,\beta_2{-}\sigma_2, \dfrac{\sigma_2}{K_1},0)$}]{} 
                        edge from parent node[right]{$D$}
                    }
                edge from parent node[left]{$F$}
            }
            child{node[hollow node,label=below:{$(0,0,0,0)$}]{} 
                edge from parent node[right]{$D$}
            }
            edge from parent node[left]{$S2$}
        }
        child{node[solid node,label=above right:{$P3$}](s3){}
            child{node[solid node,label=above left:{$P1$}]{} 
                    child{node[hollow node,label=below:{$\ \ \ \ \ \ \ \ \ \ \ (\alpha_3{+} \beta_3{-}\sigma_1,0, \dfrac{\sigma_1}{K_1},0)$}]{} 
                        edge from parent node[left]{$F$}
                    }
                    child{node[hollow node,label=right:{$(\alpha_2,\beta_2{-}\sigma_2, \dfrac{\sigma_2}{K_1},0)$}]{} 
                        edge from parent node[right]{$D$}
                    }
                edge from parent node[left]{$F$}
            }
            child{node[hollow node,label=below:{$(0,0,0,0)$}]{} 
                edge from parent node[right]{$D$}
            }
            edge from parent node[right,xshift=9.5]{$S3$}
        }; 
        \draw[dashed] (s1)to(s2)to(s3);
    \end{tikzpicture}
    \caption{Extensive form when Player 2 goes first. The optimal solution is reached when all players follow the protocol.}
\end{figure*}

%\section{Smart Meter Use Case}
%\label{use-case}
%\input{Sections/UseCase}

\section{Related Work}
\label{related-work}
There are many alternative blockchains with various properties including privacy, support for more complex smart contracts, and client software for interfacing with the blockchain, as well as with various uses of blockchain technology in IoT.

\subsection{Simplified Payment Verification}
Simplified Payment Verification (SPV) clients are lightweight Bitcoin clients \cite{BTC:whitepaper}, that do not need to store the full state of the blockchain. Instead, they store the 80-byte block headers. The block header contains a lot of information as they are chained together and contain the Merkle root of the transactions in each block. By providing an SPV client with a Merkle proof, any node can convince an SPV client that a transaction is included in the blockchain with high security as it is not efficient to create fake block headers. However, SPV clients rely on blockchain nodes to watch for payments. With many IoT devices making payments, there is no incentive for regular Bitcoin nodes to watch the blockchain for specific transactions. Therefore, we argue that SPV clients are a burden to the Bitcoin network and we design our protocol to avoid these scalability limitations. In practice, if the IoT device has storage for approximately 4 MB per year for storing block headers, then it is reasonable to include the SPV client in addition to our protocol for even greater security. However, our solution requires significantly less data storage for IoT devices.

\subsection{Payment Channels}
This work expands the Lightning Network \cite{BTC:lightning} payment channel system for IoT devices. The full Lightning Network enables payments to be routed through third parties too while we leave routing through third parties as future work in order to fully integrate with the Lightning Network. 

Bolt \cite{blockchain:bolt}, is a protocol that enables private off-chain payment channel transactions. Because of the privacy-preserving nature, it is a challenge to interface with such a system on low resource devices commonly seen in IoT. However, privacy preservation in state channels can be a desired property in IoT payment systems, such as smart meters in the power grid.

Raiden \cite{ETH:Raiden}, is an Ethereum based payment channel similar to the Lighting Network. Our protocol can also be adapted to this style network. Plasma \cite{ETH:plasma}, is a scaling solution designed on Ethereum that enables payment channels as well as more complex smart contracts to be deployed. %which may be useful for certain IoT applications. 

IOTA (MIOTA) \cite{blockchain:iota}, is a blockchain designed with low computation for IoT and web 3.0 protocols. In our case, however, the security model of this blockchain is quite different of that from Bitcoin and Ethereum. Additionally, we choose to use Bitcoin because it is the most widely used and thus easier to adopt in practice.

\subsection{IoT Integrated with Blockchain}
Christidis and Devetsikiotis explore the challenges and opportunities of blockchain and smart contracts for IoT in \cite{IoT:blockchain}. 
Their work focuses on discovering use cases that distributed ledger technology can solve and challenges found with integration of IoT. 
One challenge that they do not discuss is the resource limitations of IoT, which is the problem we propose a solution for. 
Our solution only covers the value transfer portion of blockchain technology.

In \cite{blockchain:energy:1}, Aitzhan and Svetinovic propose a token-based system similar to Bitcoin and coupled with an anonymous messaging system to provide security and privacy for peer-to-peer energy trading. 
They also include the ability to open unidirectional payment channels for partial payment. Their approach designs an anonymous market revolving around energy trading.
In our work, we focus solely on bidirectional payment channels for the Bitcoin blockchain.
While application-specific blockchains and token systems including \cite{blockchain:energy:4} may provide a solution in specific domains, we would like to explore general purpose solutions within IoT payment systems.

In \cite{blockchain:energy:3}, the authors focus on creating a local energy market for matching energy orders in a decentralized manner.
Their proposals call for a private or permissible blockchain, which operates as a decentralized trusted application. 
In this work, we focus on integrating the existing end-user devices to IoT gateways on public trustless blockchains.

In \cite{blockchain:energy:5}, Blockchain is evaluated for use in smart homes for IoT, but the blockchain proposed does not use a trustless consensus algorithm, which makes it a decentralized database for recording internet-of-things devices activity. 
Reference~\cite{blockchain:energy:8} attempts a similar objective for smart grid sensors and actuators.
In our work, we design a protocol that can be generally applied to public trustless blockchains. 
Additionally, we aim to solve the problem of value transfer rather than information assurance.

%In \cite{blockchain:energy:6}, coin where coin gen is tied to energy gen
%In \cite{blockchain:energy:7}, the authors discuss some potential issues with smart grid such as speed, scale and security and suggest that blockchain may help provide an answer.
%They call for the need of a testbed and tools to evaluate blockchain within the electric grid.
%One of the potential use cases the authors note is AMI billing which reaffirms our motivation for designing a real-time AMI billing protocol.

%In \cite{blockchain:energy:8}, the authors suggest use of a permissioned private blockchain as a distributed database for devices throughout the smart grid. 
%Their approach uses blocks to hold measurements to increase security but this approach does not scale as the measurements would need to be kept on each node in the network and their approach does not use any value tokens which prevents our real-time billing protocol from working.

Our approach uses the Bitcoin blockchain for our protocol design. There are many other blockchains that have useful properties, such as Ethereum \cite{ETH:whitepaper}, Litecoin, various Bitcoin forks, and many others which have a different block time and can implement Turing-complete scripting languages or provide differing features, such as privacy and anonymity.
However, payment channels are still in development and on-chain transactions will still suffer from the high fee problem that Bitcoin on-chain transactions do.
In our future work, we will analyze trade-offs between blockchain ecosystems for IoT and cyber-physical system payments.

\section{Conclusion}
\label{conclusion}
%The next steBitcoin payment channels are possible with
%While real time payments in advanced metering infrastructure is possible through blockchain based payment channels, 
%In our future work we would like to implement the protocol to measure the performance impact and determine hardware requirements for smart meters since they have limited processing and storage abilities.

We design a real-time blockchain-based payment channel for IoT devices to gateway services, which is less resource intensive than existing solutions, and show that off-chain payment channels are feasible for applications where IoT devices transfer value.
We also demonstrate that the protocol is crypto-economically fair by modeling the protocol as a game, in which the equilibrium is reached as long as the players follow the protocol and set the fees appropriately. 
In the future, we would like to expand our protocol for IoT devices to interact with other IoT devices as well as generalize the protocol to work with payment channels including interoperability with the existing Lightning Network \cite{BTC:lightning}. 
We would also like to explore the ability of privacy-preserving blockchain payment channels, such as Bolt \cite{blockchain:bolt}, in order to protect the rights of end-users in IoT and cyber-physical systems. 

\appendices
\section{Bitcoin TX Scripts}
\label{bitcoinScripts}
\begin{algorithm}%[h!]
    \floatname{algorithm}{Transaction}
    \caption{Funding Transaction Script OUT1}
    \label{FundTX}
    \begin{algorithmic}[1]
        %\Function{Redeem}{txIn,$\sigma$,txInRD}
        %\If{$tdf>0$}                                    %\Comment{$tsk.dilaton = 0$ before nitialization}             
        \State \textbf{Redeem Script} (r\_S)
        \Indent
            \State 2 \textless $pk_{\_A\_FT}$\textgreater \ \textless $pk_{\_B\_FT}$\textgreater\ 2 CHECKMULTISIG 
        \EndIndent
        \State \textbf{Locking Script} (l\_S)
        \Indent
            \State HASH160 \textless r\_S\_Hash\textgreater\ EQUAL
        \EndIndent
        \State \textbf{Unlocking Script} (u\_S)
        \Indent
            \State 0 \textless $sig_{\_A\_FT}$\textgreater\ \textless $sig_{\_B\_FT}$\textgreater\ \ \textless r\_S\textgreater\ \textless l\_S\textgreater\
        \EndIndent        %\EndFunction
        \State \textbf{where}
        \Indent
            \State r\_S\_Hash =      RIPEMD160 (SHA256 (r\_S))    
        \EndIndent
    \end{algorithmic}
\end{algorithm}

\begin{algorithm}%[h!]
    \floatname{algorithm}{Transaction}
    \caption{Mutual Close Transaction}
    \label{CloseTX}
    \begin{algorithmic}[1]
        %\Function{Redeem}{txIn,$\sigma$,txInRD}
        %\If{$tdf>0$}                                    %\Comment{$tsk.dilaton = 0$ before nitialization}             
        \State \textbf{In1} \Comment{FT}
        \Indent
           \State Funding Transaction (FT)
        \EndIndent
        \State \textbf{Out1} \Comment{A}
        \Indent
            \State \textbf{Locking Script} (l\_S)
            \State DUP HASH160 \textless H($pk_{\_A\_close}$)\textgreater\ 
            \Statex \hspace{12pt} EQUALVERIFY CHECKSIG
            \State \textbf{Unlocking Script} (u\_S)
            \State \textless $sig_{\_A\_close}$\textgreater\ \textless $pk_{\_A\_close}$\textgreater\
        \EndIndent
        \State \textbf{Out2} \Comment{B}
        \Indent
            \State \textbf{Locking Script} (l\_S)
            \State DUP HASH160 \textless H($pk_{\_B\_close}$)\textgreater\ \Statex \hspace{12pt} EQUALVERIFY CHECKSIG
            \State \textbf{Unlocking Script} (u\_S)
            \State \textless $sig_{\_B\_close}$\textgreater\ \textless $pk_{\_B\_close}$\textgreater\
        \EndIndent     %\EndFunction
        \State \textbf{where}
        \Indent
            \State H($pk$) = RIPEMD160 (SHA256 ($pk$))%\footnotemark  
        \EndIndent
    \end{algorithmic}
\end{algorithm}

%\todo{define Q,K}

\begin{algorithm*}%[h!]
    \floatname{algorithm}{Transaction}
    \caption{Commitment $i\_a$ (Publishable by A)}
    \label{commitmentTXa}
    \begin{algorithmic}[1]
        %\Function{Redeem}{txIn,$\sigma$,txInRD}
        %\If{$tdf>0$}                                    %\Comment{$tsk.dilaton = 0$ before nitialization}             
        \State \textbf{In1} \Comment{FT}
        \Indent
           \State Funding Transaction (FT)
        \EndIndent
        \State \textbf{Out1} \Comment{A}
        \Indent
            \State \textbf{Locking Script} (l\_S)
            \State IF 
            \Indent 
                \State W CHECKSEQUENCEVERIFY DROP DUP HASH160 \textless H($pk_{\_A\_i\_a}$)\textgreater\ EQUALVERIFY CHECKSIG
            \EndIndent
            \State ELSE
            \Indent
            \State DUP HASH160 \textless H($pk_{\_B\_i\_b}$)\textgreater\ EQUALVERIFY CHECKSIG DROP
            \Statex \hspace{27pt} DUP HASH160 \textless H($pk_{\_A\_i\_c}$)\textgreater\ EQUALVERIFY CHECKSIG
            \EndIndent
            \State ENDIF
            %\State \textbf{scriptPubKey:}
            %\State DUP HASH160 \textless H($pk_{\_A\_close}$)\textgreater\ EQUALVERIFY CHECKSIG
            %\State \textbf{scriptSig:}
            %\State \textless $sig_{\_A\_close}$\textgreater\ \textless $pk_{\_A\_close}$\textgreater\
        \EndIndent
        \State \textbf{Out2} \Comment{B}
        \Indent
            \State \textbf{Locking Script} (l\_S)
            \State DUP HASH160 \textless H($pk_{\_B\_i\_b}$)\textgreater\ EQUALVERIFY CHECKSIG
            %\State \textbf{Unlocking Script} (u\_S)
            %\State \textless $sig_{\_B\_close}$\textgreater\ \textless $pk_{\_B\_close}$\textgreater\
        \EndIndent     %\EndFunction
        \State \textbf{Out3} \Comment{3rd Party A (Economically Incentivized to publish to the blockchain)}
        \Indent
            \State 1 \textless $pk_{\_3rd\_a\_0}$\textgreater \ \textless $pk_{\_3rd\_a\_K}$\textgreater\ $K_1$ CHECKMULTISIG 
            %\State \textbf{Unlocking Script} (u\_S)
            %\State \textless $sig_{\_B\_close}$\textgreater\ \textless $pk_{\_B\_close}$\textgreater\
        \EndIndent     %\EndFunction
    \end{algorithmic}
\end{algorithm*}

\begin{algorithm*}%[h!]
    \floatname{algorithm}{Transaction}
    \caption{Commitment $i\_b$ (Publishable by B)}
    \label{commitmentTXb}
    \begin{algorithmic}[1]
        %\Function{Redeem}{txIn,$\sigma$,txInRD}
        %\If{$tdf>0$}                                    %\Comment{$tsk.dilaton = 0$ before nitialization}             
        \State \textbf{In1} \Comment{FT}
        \Indent
           \State Funding Transaction (FT)
        \EndIndent
        \State \textbf{Out1} \Comment{B}
        \Indent
            \State \textbf{Locking Script} (l\_S)
            \State IF 
            \Indent 
                \State W CHECKSEQUENCEVERIFY DROP DUP HASH160 \textless H($pk_{\_B\_i\_a}$)\textgreater\ EQUALVERIFY CHECKSIG
            \EndIndent
            \State ELSE \Comment{3rd Party B (Economically Incentivized to watch the blockchain)}
            \Indent
            \State 1 \textless $pk_{\_3rd\_b\_0}$\textgreater \ \textless $pk_{\_3rd\_b\_K}$\textgreater\ $K_2$ CHECKMULTISIG DROP
            \Statex \hspace{27pt}  DUP HASH160  \textless H($pk_{\_B\_i\_c}$)\textgreater\ EQUALVERIFY CHECKSIG DROP
            \Statex \hspace{27pt} DUP HASH160  \textless H($pk_{\_A\_i\_b}$)\textgreater\ EQUALVERIFY CHECKSIG 
            \EndIndent
            \State ENDIF
            %\State \textbf{scriptPubKey:}
            %\State DUP HASH160 \textless H($pk_{\_A\_close}$)\textgreater\ EQUALVERIFY CHECKSIG
            %\State \textbf{scriptSig:}
            %\State \textless $sig_{\_A\_close}$\textgreater\ \textless $pk_{\_A\_close}$\textgreater\
        \EndIndent
        \State \textbf{Out2} \Comment{A}
        \Indent
            \State \textbf{Locking Script} (l\_S) 
            \State DUP HASH160  \textless H($pk_{\_A\_i\_b}$)\textgreater\ EQUALVERIFY CHECKSIG
            %\State \textbf{Unlocking Script} (u\_S)
            %\State \textless $sig_{\_B\_close}$\textgreater\ \textless $pk_{\_B\_close}$\textgreater\
        \EndIndent     %\EndFunction
    \end{algorithmic}
\end{algorithm*}

\begin{algorithm*}%[h!]
\floatname{algorithm}{Transaction}
    \caption{Recovery Transaction}
    \label{RecoverTX}
    \begin{algorithmic}[1]
        %\Function{Redeem}{txIn,$\sigma$,txInRD}
        %\If{$tdf>0$}                                    %\Comment{$tsk.dilaton = 0$ before nitialization}             
        \State \textbf{In1} \Comment{Commitment TX $b$}
        \Indent
            \State  (Transaction \ref{commitmentTXb}, Output 1)
            \State \textbf{Unlocking Script} (u\_S)
            \State \textless $sig_{\_A\_i\_b}$\textgreater\ \textless $pk_{\_A\_i\_b}$\textgreater\ \textless $sig_{\_B\_i\_c}$\textgreater\  \textless $pk_{\_B\_i\_c}$\textgreater\  \textless $pk_{\_3rd\_b\_\alpha}$\textgreater\
        \EndIndent
        \State \textbf{Out1} \Comment{A}
        \Indent
            \State \textbf{Locking Script} (l\_S)
            \State DUP HASH160 \textless H($pk_{\_A\_i\_rc}$)\textgreater\  EQUALVERIFY CHECKSIG
            \State \textbf{Unlocking Script} (u\_S)
            \State \textless $sig_{\_A\_i\_rc}$\textgreater\ \textless $pk_{\_A\_i\_rc}$\textgreater\
        \EndIndent
        \State \textbf{Out2} \Comment{3rd Party b}
        \Indent
            \State \textbf{Locking Script} (l\_S)
            \State DUP HASH160 \textless H($pk_{\_3rd\_b\_\omega}$)\textgreater\  EQUALVERIFY CHECKSIG
        \EndIndent
        \State \textbf{Out3} \Comment{3rd Party a}
        \Indent
            \State \textbf{Locking Script} (l\_S)
            \State 1 \textless $pk_{\_3rd\_a\_0}$\textgreater \ \textless $pk_{\_3rd\_a\_K}$\textgreater\ $K_1$ CHECKMULTISIG DROP
        \EndIndent
    \end{algorithmic}
\end{algorithm*}

%\footnotetext{Technically pubKeyHashes also include version, and checksum in base58 encoding.}  

%\section{}

%\section*{Acknowledgment}

%The authors would like to thank Anita Nikolich, Yuan Hong, and Mohit Hota for their advice and suggestions.
%\todo{add funding? we can add this part later once the paper is accepted}

\ifCLASSOPTIONcaptionsoff
  \newpage
\fi



%\bibliographystyle{IEEEtran}
%\begin{thebibliography}{1}


%\end{thebibliography}
\bibliography{main} 
\bibliographystyle{ieeetr}
\end{document}